# Automated family-based naming of small RNAs for next generation sequencing data using a modified MD5-digest algorithm

Guodong Liu[1], Zhihua Li[1], Yuefeng Lin[1], and Bino John[1,2]

[1]Department of Computational Biology, University of Pittsburgh School of Medicine, Pittsburgh, PA 15260, USA

[2]Molecular Virology Program, Hillman Cancer Center, 5117 Centre Avenue, Pittsburgh, PA 15213, USA

Corresponding author:

Bino John

University of Pittsburgh School of Medicine

3080 Biomedical Science Tower 3

3501 Fifth Avenue

Pittsburgh, PA 15260, USA

Phone: 412-648-8607; Fax: 412-648-3163

Email: john@johnlab.org

Web: http://www.johnlab.org

October 23, 2012

## Abstract

**Summary:** We developed NameMyGene, a web tool and a stand alone program to easily generate putative family-based names for small RNA sequences so that laboratories can easily organize, analyze, and observe patterns from, the massive amount of data generated by next-generation sequencers. NameMyGene, also applicable to other emerging methods such as RNA-Seq, and Chip-Seq, solely uses the input small RNA sequence and does not require any additional data such as other sequence data sets.

**Availability and Implementation**: The web server and software is freely available (http://www.johnlab.org/NameMyGene) and is based on Java to ensure platform independency.

**Contact:** john@johnlab.org

**Supplementary information**: Online-only Supplementary data is available at the journal's web site.

## Introduction

Several distinct classes of small RNAs (20–30 bases) such as microRNAs (miRNAs), play important biological roles in diverse cellular processes across a wide range of species (Ghildiyal et al.,2009). In addition to the well characterized miRNAs, other important small RNAs include endogenous small interfering RNAs (endo-siRNAs), piwi-interacting RNAs (piRNAs), and transcription start-site associated RNAs (TSSa-RNAs). With rapid advances in next generation sequencing techniques (Shendure et al.,2008), new small RNAs continue to get discovered at an unprecedented rate.

A key step in annotating small RNAs from the massive amount of data generated by next generation sequencers is the assignment of unique names to small RNAs and providing a putative familial classification (putative homologs) for them. However, small RNAs identified using sequencing approaches are arbitrarily named by individual laboratories and their names frequently do not reflect their evolutionary relationships, raising considerable difficulties in organizing and analyzing these sequences, and recognizing their sequence characteristics during planning experiments. For instance, cross-hybridization of detection probes of target RNAs to irrelevant sequences could be reduced, if putative homologs were considered during probe design (*e.g.*, LNA spiking patterns (Varallyay et al.,2008)). Similarly, miRNA target validation experiments benefit from the knowledge of whether a specific family of miRNA sequences targets a given gene.

The aforementioned drawbacks of sequence names are arguably minimal for the well-established class of miRNAs, which has reasonably well defined rules on sequence annotation (Ambros et al.,2003). However, even for miRNAs such as the widely known miR-17 family members (*e.g.*, miR-93 and -106), we frequently cannot readily infer their evolutionary relationships. However, providing a simple uniform naming system to reflect evolutionary relationships is a hard problem. Even if we adopt a nomenclature where all family members are named analogously, a newly found small RNA must be compared to all existing small RNAs from one or more massive sequence data sets to assign it to the correct family, a task requiring significant bioinformatics support that is unavailable to many laboratories. Thus, a sequence naming system that relies exclusively on the sequence of interest and automatically provide the putative family of the sequence is highly desirable and helpful to many biologists in their laboratories. Such RNA names are also easy to communicate, understand, or remember in scientific presentations particularly when novel small RNA sequences are presented. Therefore, to help generate automatic names, yet a useful putative family classification of small

RNAs, we developed NameMyGene (http://www.johnlab.org/NameMyGene), a free web service that can automatically assign unique gene names and families to small RNAs. NameMyGene is also freely available for download as open source software, implemented in Java for platform independency.

## Methods

### Structure of small RNA names

NameMyGene generates small RNA names containing a sequence-based field and a familial signature, separated by a hyphen. For example, Kdo94-H2V represents sequence Kdo94 of the H2V sequence family. The first field (*e.g.*, Kdo94) could be used as a common name of the RNA gene and is generated using a modified (**Doc-S1**) version of the Message Digest (Rivest ,1992) Algorithm (MD5), termed compact MD5 (CMD5). The use of CMD5 ensures that the name is probabilistically unique to the given sequence. The second field, the familial signature, is designed to quickly infer whether two or more sequences share sequence similarity.

### CMD5 algorithm

To generate mnemonically friendly names, we chose to create names that contain 3 consecutive alphabets (A–Z) followed by a maximum of two letters (0–9), thus allowing for a large number of 1,757,600 ($26^3 \times 10^2$) distinct, yet mnemonically names. We therefore built upon MD5 (Rivest ,1992), a widely used algorithm for message encryption in cryptography. The MD5 output is a 128-bit cryptic message that is long and therefore is not suitable for assigning short names to RNA sequences. We therefore developed CMD5 (**Doc-S**), optimized for encrypting small RNA sequences (< 40 nts). Briefly, CMD5 first breaks the binary sequence representation of RNAs into 96-bit blocks. Analogous to MD5, CMD5 processes each block in a step-by-step manner to continuously update a 24-bit process state. The processing of a sequence block consists of three message processing stages, followed by a final stage that is unique to CMD5. To obtain the RNA name, CMD5 converts the 24-bit state into a final output string consisting of three letters chosen from the 26-character alphabet set (A–Z) and two digits (0–9) from the decimal system. The output string is obtained by first converting the 24-bit sequence to the decimal system and performing recursive divisions on the integer (**Doc-S**). The resulting three alphabetic characters and two digits are concatenated to yield the CMD5-based identifier for a given RNA sequence.

**Generation of familial signature based on optimal leaf ordering of hexamers**

The three character familial signature is derived based on the three hexamer sequences present at positions 2–9 of a given small RNA sequence. Each hexamer is mapped to one of the 26 letters (A–Z) that encodes a pre-constructed sequence library containing the hexamer and another ~157 closely related hexamers. The 26 libraries are constructed by average linkage-based hierarchical clustering and optimal leaf ordering (Bar-Joseph et al.,2001) of all (4,096) possible hexamers, based on their sequence identities determined using local dynamic programming (MATLAB). The three letters that map to the three hexamers are concatenated together to yield the final familial signature. For mnemonic simplicity, if a given letter is tandemly repeated (*e.g.*, V in HVV), the repeating letters are replaced by the letter followed by its number of occurrences (*e.g.*, HVV by HV2).

## Results and Discussion

We tested our method using 4,858 miRNA sequences that were classified by miRBase (Griffiths-Jones et al.,2008) into 583 families. The average percentage of RNAs classified by NameMyGene over all 583 families is 88.3% (**Table S1**). For instance, all human let-7 family members are accurately classified into the H2V-family of sequences and the majority (6/8) of the human miR-17 family of sequences is categorized as NI2 (**Table 1**). Notably, the miR-18a/b members of the miR-17 family are represented by HI2, a signature that closely resembles the NI2 signature. Thus, in addition to representing very closely related sequences by a single familial signature, the proposed method also accurately captures the close similarity between small RNAs as observed for the miR-18a/b sub-family and other more distant members of the miR-17 family.

Our method does not use any additional information other than the sequence of interest, and therefore provides a quick and useful means to organize newly identified small RNAs across different small RNA classes and species. NameMyGene is also useful for organizing sequences and in identifying patterns in short sequences associated with other high throughput methods such as RNA-Seq, Chip-Seq, SAGE, and genome tiling arrays (*e.g.*, probe names). Finally, the CMD5 generated fields of NameMyGene could also be useful to generate mnemonically friendly names for any DNA or RNA sequence of interest.

## Acknowledgments

We thank laboratory members for helpful discussions, Mr. Boles for computer systems administration. We also acknowledge the grant supports provided by the University of Pittsburgh, and NIGMS/NIH (R01GM079756).

| Species | miRNA | miRNA sequence | Automated Name | Familial Signature |
|---|---|---|---|---|
| Human | miR-106a | AAAAGUGCUUACAGUGCAGGUAG | Gqd85 | NI2 |
| | miR-106b | UAAAGUGCUGACAGUGCAGAU | Vik36 | NI2 |
| | miR-17 | CAAAGUGCUUACAGUGCAGGUAG | Tuq99 | NI2 |
| | miR-20a | UAAAGUGCUUAUAGUGCAGGUAG | Wxn52 | NI2 |
| | miR-20b | CAAAGUGCUCAUAGUGCAGGUAG | Tmu61 | NI2 |
| | miR-93 | CAAAGUGCUGUUCGUGCAGGUAG | Voq38 | NI2 |
| | miR-18a | UAAGGUGCAUCUAGUGCAGAUAG | Ujr2 | HI2 |
| | miR-18b | UAAGGUGCAUCUAGUGCAGUUAG | Yql70 | HI2 |
| Mouse | miR-106a | CAAAGUGCUAACAGUGCAGGUAG | Ypn14 | NI2 |
| | miR-106b | UAAAGUGCUGACAGUGCAGAU | Vik36 | NI2 |
| | miR-17 | CAAAGUGCUUACAGUGCAGGUAG | Tuq99 | NI2 |
| | miR-20a | UAAAGUGCUUAUAGUGCAGGUAG | Wxn52 | NI2 |
| | miR-20b | CAAAGUGCUCAUAGUGCAGGUAG | Tmu61 | NI2 |
| | miR-93 | CAAAGUGCUGUUCGUGCAGGUAG | Voq38 | NI2 |
| | miR-18a | UAAGGUGCAUCUAGUGCAGAUAG | Ujr2 | HI2 |
| | miR-18b | UAAGGUGCAUCUAGUGCUGUUAG | Voq68 | HI2 |

**Table 1: Automatically generated names for the human and mouse miR-17 sequence family**. The sequence similarities between all miR-17 family members could be readily identified using their sequence signature (*e.g.*, NI2 for miR-106a and miR-20b), while the automated names provide a means to distinguish between sequence variants (*e.g.*, human miR-106a and mouse miR-106a). The familial signatures and the sequence-dependent automated names are shown to illustrate the use of the proposed method for naming novel, poorly understood small RNAs and are not meant to replace miRNA names.

## References

Ambros, V., Bartel, B., Bartel, D. P., Burge, C. B., Carrington, J. C., Chen, X., Dreyfuss, G., Eddy, S. R., Griffiths-Jones, S., Marshall, M., Matzke, M., Ruvkun, G., and Tuschl, T. (2003). A uniform system for microRNA annotation. RNA. 9, 277-279.

Bar-Joseph, Z., Gifford, D. K., and Jaakkola, T. S. (2001). Fast optimal leaf ordering for hierarchical clustering. Bioinformatics 17 Suppl 1, S22-S29.

Ghildiyal, M. and Zamore, P. D. (2009). Small silencing RNAs: an expanding universe. Nat.Rev.Genet. 10, 94-108.

Griffiths-Jones, S., Saini, H. K., van, Dongen S., and Enright, A. J. (2008). miRBase: tools for microRNA genomics. Nucleic Acids Res. 36, D154-D158.

Rivest, R. L. The MD5 message-digest algorithm. RFC 1321 37.  1992.

Shendure, J. and Ji, H. (2008). Next-generation DNA sequencing. Nat.Biotechnol. 26, 1135-1145.

Varallyay, E., Burgyan, J., and Havelda, Z. (2008). MicroRNA detection by northern blotting using locked nucleic acid probes. Nat.Protoc. 3, 190-196.

**Supplementary information**

In the original MD5 algorithm, a given message is processed into a fixed-length output of 128 bits. Messages are generally encoded *in silico* using binaries of the ASCII values of their characters. For instance, the di-nucleotide sequence, AC, can be encoded by a 12 (2 x 6) bit binary (010000010100). Our algorithm, CMD5, uses a three stage process that resembles the core MD5 algorithm to process the binary sequence of a given RNA (**Stages 1–3**).

**Stage 1: Append bits to the binary sequences**. Given integers *a*, *b*, and *n*, we consider *a* congruent to *b*, modulo (mod) *n*, if *b* is the remainder when *a* is divided by *n*. CMD5 first appends "1" followed by "0"s to the end of the binary sequence, such that the bit-length of the padded sequence is congruent to 80, modulo 96. In addition, the bit-length of the sequence prior to padding is converted to a 16-bit binary sequence, and the resulting 16-bit sequence is appended to the end of the padded sequence.

**Stage 2: Initialize state.** CMD5 uses four 6-bit words (*a*, *b*, *c*, and *d*) to store a 24-bit (6 x 4) process state. The four binary words are initialized as: 000000 (*a*), 010000 (*b*), 100000 (*c*), and 110000 (*d*).

**Stage 3: Encode a binary RNA sequence.** Four logical functions (F, G, H, and I) perform a series of logical bitwise operations, on a given set of three inputs (X, Y, and Z), based on four basic operators (**NOT, AND**, **OR**, and **XOR)**.

*F(X,Y,Z) = (X **AND** Y) **OR** (**NOT** (X) **AND** Z)*
*G(X,Y,Z) = (X **AND** Z) **OR** (Y **NOT** (Z))*
*H(X,Y,Z) = X **XOR** Y **XOR** Z*
*I(X,Y,Z) = Y **XOR** (X **OR** (**NOT** (Z)))*

Briefly, the four operators perform bit-wise operations such that the $i^{th}$ bit of the operation output is solely determined by the $i^{th}$ bits of the inputs, independent from the values of other bits (**Table 1**). **NOT** operates on a single input to convert each input bit into its complement (1 to 0, and 0 to 1). **AND, OR**, and **XOR** operate on two input binary sequences of equal length and operate on the pairs of the corresponding bits. In each pair, **AND** returns 1, if and only if both bits are 1, and returns 0 otherwise; **OR** returns 1, if and only if either or both bits are 1, and returns 0 otherwise; **XOR** returns 1, if and only if the bits are different from each other, and returns 0 otherwise. Thus, the output of each function is a 6-bit binary sequence (**Table 2**).

| Input/Output | Logical Operations | | | |
|---|---|---|---|---|
| | NOT | OR | AND | XOR |
| Input 1 | 010110 | 010110 | 010110 | 010110 |
| Input 2 | NA | 011011 | 011011 | 011011 |
| Output | 101001 | 011111 | 010010 | 001101 |

Table 1: Illustration of bit-wise operations by NOT, OR, AND, and XOR.

The four logical functions, an operator termed "left circular shifting", and another function termed "modular addition", operate on the sixteen 6-bit sub-blocks of the 96-bit block to update the 24-bit state value. The left circular shifting operator is used to transpose bits from the left end to the right end. For example, left circular shifting 011001 by two bits yields 100101. Modular addition is similar to numerical addition. The operator processes two binary sequences as inputs, adding two bits at a time, from right to left. Any carryover beyond the leftmost bit is ignored. For example, modular addition of 011001 and 110101 produces 001110.

| Steps | Expressions | Expression Values |
|---|---|---|
| 1 | X AND Y | 010000 |
| 2 | NOT (X) | 101011 |
| 3 | NOT (X) AND Z | 100001 |
| 4 | (X AND Y) OR (NOT (X) AND Z) | 110001 |

**Table 2: An example of a step-by-step operation of the logical function F.** Parameters used are: X=010100, Y=011011, and Z=110101.

Stage 3 consists of 64 sets of operations. The operations are divided into 4 rounds, with 16 sets of operations in each round. Each round applies a single function repeatedly, 16 times. CMD5 maintains a 64-element auxiliary array T[1,2, … 64], the elements are progressively used in each set of operations. T[$i$] represents the integer part of 64×abs(sin($i$)), used in the $i^{th}$ set of operations. The four functions are:

*Function Round1 (a, b, c, d, xid, tid, sb)*
   *a = b + ((a + LeftCircularShift(F(b,c,d) + X[xid] +T[tid]), sb));*
*Function Round2 (a, b, c, d, xid, tid, sb)*
   *a = b + ((a + LeftCircularShift(G(b,c,d) + X[xid] +T[tid]), sb));*
*Function Round3 (a, b, c, d, xid, tid, sb)*
   *a = b + ((a + LeftCircularShift(H(b,c,d) + X[xid] +T[tid]), sb));*
*Function Round4 (a, b, c, d, xid, tid, sb)*
   *a = b + ((a + LeftCircularShift(I(b,c,d) + X[xid] +T[tid]), sb));*

*xid* corresponds to the index of one of the 16 words at the current block, *X* (see pseudocode), *tid* corresponds to one of the 64 elements in array T, and *sb* represents the number of bits to be shifted. The parameters (**Table 3**) passed to all 64 sets of operations are identical to the MD5 algorithm. The pseudocode for stage 3 is:

*For i = 1 to NumberOfBlocks do*
   *Assign 96-bit long block [i] to X[0…15], such that X[xid] contains a six-bit sequence*
   *Save the process state values a, b, c, and d to aa, bb, cc, and dd*
   *Execute 64 sets of operations, from round 1 to round 4*

   *a = a + aa;*
   *b = b + bb;*
   *c = c + cc;*
   *d = d + dd;*

*End of loop*
 *FinalState [0:5] = a [0:5];*
 *FinalState [6:11] = b [0:5];*
 *FinalState [12:17] = c [0:5];*
 *FinalState [18:23] = d [0:5];*

Finally, the output string is obtained by first converting the 24-bit sequence to the decimal system and performing recursive divisions on the integer, as illustrated in the pseudocode below:

*Number = 0;*
*For i = 0 to 23 do*
     *Number = Number + FinalState [i] × $2^i$;*
*End of loop*
*Remainder1  = remainder ( Number, 26×26×26×10×10);*
*Letter1_Index = (int) Remainder1 / (26×26×10×10);*
*Remainder2  = remainder ( Remainder1, 26×26×10×10);*
*Letter2_Index = (int) Remainder2 / (26×10×10);*
*Remainder3  = remainder ( Remainder2, 26×10×10);*

*Letter3_Index = (int) Remainder3 / (10×10);*
*Remainder4 = remainder ( Remainder3, 10);*
*Digit1 = (int) Remainder / (10);*
*Remainder5 = remainder ( Remainder4, 10);*
*Digit2 = Remainder5;*
*Letter1 = Alphabet [Letter1_Index] ;*
*Letter2 = Alphabet [Letter2_Index];*
*Letter3 = Alphabet [Letter3_Index] ;*
*FinalCode = concatenate (Letter1, Letter2, Letter3, Digit1, Digit2);*
*Output FinalCode;*

| Operation Set | Round 1 | Round 2 | Round3 | Round4 |
|---|---|---|---|---|
| 1 | *a,b,c,d*,0,7,1 | *a,b,c,d*,1,5,17 | *a,b,c,d*,5,4,33 | *a,b,c,d*,0,6,49 |
| 2 | *d,a,b,c*,1,12,2 | *d,a,b,c*,6,9,18 | *d,a,b,c*,8,11,34 | *d,a,b,c*,7,10,50 |
| 3 | *c,d,a,b*,2,17,3 | *c,d,a,b*,11,14,19 | *c,d,a,b*,11,16,35 | *c,d,a,b*,14,15,51 |
| 4 | *b,c,d,a*,3,22,4 | *b,c,d,a*,0,20,20 | *b,c,d,a*,14,23,36 | *b,c,d,a*,5,21,52 |
| 5 | *a,b,c,d*,4,7,5 | *a,b,c,d*,5,5,21 | *a,b,c,d*,1,4,37 | *a,b,c,d*,12,6,53 |
| 6 | *d,a,b,c*,5,12,6 | *d,a,b,c*,10,9,22 | *d,a,b,c*,4,11,38 | *d,a,b,c*,3,10,54 |
| 7 | *c,d,a,b*,6,17,7 | *c,d,a,b*,15,14,23 | *c,d,a,b*,7,16,39 | *c,d,a,b*,10,15,55 |
| 8 | *b,c,d,a*,7,22,8 | *b,c,d,a*,4,20,24 | *b,c,d,a*,10,23,40 | *b,c,d,a*,1,21,56 |
| 9 | *a,b,c,d*,8,7,9 | *a,b,c,d*,9,5,25 | *a,b,c,d*,13,4,41 | *a,b,c,d*,8,6,57 |
| 10 | *d,a,b,c*,9,12,10 | *d,a,b,c*,14,9,26 | *d,a,b,c*,0,11,42 | *d,a,b,c*,15,10,58 |
| 11 | *c,d,a,b*,10,17,11 | *c,d,a,b*,3,14,27 | *c,d,a,b*,3,16,43 | *c,d,a,b*,6,15,59 |
| 12 | *b,c,d,a*,11,22,12 | *b,c,d,a*,8,20,28 | *b,c,d,a*,6,23,44 | *b,c,d,a*,13,21,60 |
| 13 | *a,b,c,d*,12,7,13 | *a,b,c,d*,13,5,29 | *a,b,c,d*,9,4,45 | *a,b,c,d*,4,6,61 |
| 14 | *d,a,b,c*,13,12,14 | *d,a,b,c*,2,9,30 | *d,a,b,c*,12,11,46 | *d,a,b,c*,11,10,62 |
| 15 | *c,d,a,b*,14,17,15 | *c,d,a,b*,7,14,31 | *c,d,a,b*,15,16,47 | *c,d,a,b*,2,15,63 |
| 16 | *b,c,d,a*,15,22,16 | *b,c,d,a*,12,20,32 | *b,c,d,a*,2,23,48 | *b,c,d,a*,9,21,64 |

**Table 3: Parameters passed in the four rounds.** The sixteen sets of parameters are first used in round 1, followed by rounds 2, 3, and 4.